\documentclass[a4paper,aps,twocolumn,nofootinbib]{revtex4}
\RequirePackage[colorlinks,hyperindex]{hyperref}
\RequirePackage[english]{babel}
\RequirePackage[latin1]{inputenc}
\RequirePackage[T1]{fontenc}
\RequirePackage{mathrsfs}
\RequirePackage{amsmath}
\RequirePackage{amssymb}
\RequirePackage{amsbsy}
\RequirePackage{color}
\RequirePackage{bm}
\hypersetup{colorlinks=true,breaklinks=true,urlcolor=blue,linkcolor=red}
\begin{document}
\title{\bf{Covariant Inertial Forces for Spinors}}
\author{Luca Fabbri}
\affiliation{DIME, Universit\`{a} di Genova,\\ P.Kennedy Pad.D, 16129 Genova, ITALY}
\date{\today}
\begin{abstract}
In this paper we consider the Dirac spinor field in interaction with a background of electrodynamics and torsion-gravity; by performing the polar reduction we acquire the possibility to introduce a new set of objects that have the geometrical status of non-vanishing tensors but which seem to contain the same information of the connection: thus they appear to be describing something that seems like an inertial force but which is also essentially covariant. After a general introduction, we exemplify these tensors in the very well known instance of the orbital of minimal energy for an electron in the case of the Hydrogen atom: we will see that the invariants built with these tensors remain different from zero even for free field configurations. An outlook regarding possible interpretations of such a set of tensors will be sketched. A few final comments will eventually be given.
\end{abstract}
\maketitle
\section{Introduction}
As it is well known, Dirac spinor fields can be classified in terms of the so-called Lounesto classification according to two classes: singular spinor fields are those subject to the conditions $i\overline{\psi}\boldsymbol{\pi}\psi\!\equiv\!0$ and $\overline{\psi}\psi\!\equiv\!0$ while regular spinor fields are all those for which the two above conditions do not identically hold \cite{L,Cavalcanti:2014wia,Fabbri:2016msm}; for regular spinor fields, it is possible to perform what is known as polar decomposition of the Dirac spinor field and from which it is also possible to have the polar decomposition of Dirac spinor field equations \cite{h1,Fabbri:2016laz}. The polar decomposition of Dirac spinor field equations is essentially the way in which these four complex spinor field equations can be converted into two real vectorial field equations specifying all derivatives of the two real degrees of freedom of the Dirac spinor field as explained in \cite{Fabbri:2016msm,Fabbri:2016laz}. For a broader review see also \cite{Fabbri:2017fac}.

In \cite{Fabbri:2017pwp} we have for the first time pointed out that, when the polar reduction is performed, some background quantities arise, which can be used, with the spin connection, to construct objects of a very specific type: these objects are built only in terms of entities arising from local frame and general coordinate transformations, and as such they should behave like connections, but nonetheless they have been proven to be tensors. In \cite{Fabbri:2017xlx} we have discussed these tensors in one specific case involving electrodynamics and torsion-gravity. In this paper we deepen the discussion.
\section{Geometry of the Dirac Spinor}
In the introduction we recalled the basic concepts, and now we will give them a proper mathematical definition.

\subsection{Kinematic general quantities: polar spinor}
To begin, we recall that $\boldsymbol{\gamma}^{a}$ are Clifford matrices, from which $\left[\boldsymbol{\gamma}_{a}\!,\!\boldsymbol{\gamma}_{b}\right]\!=\!4\boldsymbol{\sigma}_{ab}$ and $2i\boldsymbol{\sigma}_{ab}\!=\!\varepsilon_{abcd}\boldsymbol{\pi}\boldsymbol{\sigma}^{cd}$ are the definitions of the $\boldsymbol{\sigma}_{ab}$ and the $\boldsymbol{\pi}$ matrix (this matrix is what is usually indicated as gamma with an index five, but since in the space-time this index has no meaning we prefer to employ a notation in which no index appears at all).

For regular spinor fields, as already discussed \cite{Fabbri:2016msm,Fabbri:2016laz}, it is always possible to write them according to the expression
\begin{eqnarray}
&\!\psi\!=\!\phi e^{-\frac{i}{2}\beta\boldsymbol{\pi}}
\boldsymbol{S}\left(\!\begin{tabular}{c}
$1$\\
$0$\\
$1$\\
$0$
\end{tabular}\!\right)
\label{spinor}
\end{eqnarray}
up to the $\psi'\!=\!\boldsymbol{\pi}\psi$ transformation and up to the reversal of the third axis, with $\boldsymbol{S}$ being a generic complex Lorentz transformation: this is the polar form of the spinor field.

\subsection{Kinematic general quantities: bilinear quantities}
With $\overline{\psi}\!=\!\psi^{\dagger}\boldsymbol{\gamma}_{0}$ we indicate the operation of spinor field conjugation, from which we define the bilinear quantity
\begin{eqnarray}
&M_{ab}\!=\!2i\overline{\psi}\boldsymbol{\sigma}_{ab}\psi
\!=\!2\phi^{2}(\cos{\beta}u^{j}s^{k}\varepsilon_{jkab}\!+\!\sin{\beta}u_{[a}s_{b]})
\end{eqnarray}
in terms of independent bilinear axial-vector and vector
\begin{eqnarray}
&S^{a}\!=\!\overline{\psi}\boldsymbol{\gamma}^{a}\boldsymbol{\pi}\psi\!=\!2\phi^{2}s^{a}\\
&U^{a}\!=\!\overline{\psi}\boldsymbol{\gamma}^{a}\psi\!=\!2\phi^{2}u^{a}
\end{eqnarray}
and the bilinear pseudo-scalar and scalar
\begin{eqnarray}
&\Theta\!=\!i\overline{\psi}\boldsymbol{\pi}\psi\!=\!2\phi^{2}\sin{\beta}\label{b2}\\
&\Phi\!=\!\overline{\psi}\psi\!=\!2\phi^{2}\cos{\beta}\label{b1}
\end{eqnarray}
with $u^{a}$ the velocity vector and $s^{a}$ the spin axial-vector and where $\phi$ is a scalar and $\beta$ is a pseudo-scalar known as module and Yvon-Takabayashi angle (we stress that the name Takabayashi can sometimes be spelled Takabayasi).

One can easily prove that the directions are such that
\begin{eqnarray}
&u_{a}u^{a}\!=\!-s_{a}s^{a}\!=\!1\\
&u_{a}s^{a}\!=\!0
\end{eqnarray}
and we notice that the velocity vector has only the temporal component while the spin axial-vector has only the third component when the polar form (\ref{spinor}) is taken in the case in which $\boldsymbol{S}$ is the identity: therefore the module and the Yvon-Takabayashi angle are the only two real degrees of freedom as is most manifest when the polar form with choice $\boldsymbol{S}\!=\!\mathbb{I}$ is taken into account for the spinorial field.

\subsection{Dynamic field equations: background structure}
With $\Omega^{a}_{b\pi}$ spin connection and $A_{\mu}$ gauge potential, we can next define the space-time and gauge curvatures as
\begin{eqnarray}
&R^{i}_{\phantom{i}j\mu\nu}\!=\!\partial_{\mu}\Omega^{i}_{\phantom{i}j\nu}
\!-\!\partial_{\nu}\Omega^{i}_{\phantom{i}j\mu}
\!+\!\Omega^{i}_{\phantom{i}k\mu}\Omega^{k}_{\phantom{k}j\nu}
\!-\!\Omega^{i}_{\phantom{i}k\nu}\Omega^{k}_{\phantom{k}j\mu}\\
&F_{\mu\nu}\!=\!\partial_{\mu}A_{\nu}\!-\!\partial_{\nu}A_{\mu}
\end{eqnarray}
and where for the moment all is in the torsionless case.

If we proceed to calculate the derivative of the spinorial field in polar form (\ref{spinor}) and because for the most general complex Lorentz transformation we can always write 
\begin{eqnarray}
&\boldsymbol{S}\partial_{\mu}\boldsymbol{S}^{-1}\!=\!i\partial_{\mu}\theta\mathbb{I}
\!+\!\frac{1}{2}\partial_{\mu}\theta_{ij}\boldsymbol{\sigma}^{ij}\label{Lorentz}
\end{eqnarray}
where $\theta$ is a generic complex phase and $\theta_{ij}\!=\!-\theta_{ji}$ are the six parameters of the Lorentz group, then we can define
\begin{eqnarray}
&\partial_{\mu}\theta_{ij}\!-\!\Omega_{ij\mu}\!\equiv\!R_{ij\mu}\label{R}\\
&\partial_{\mu}\theta\!-\!qA_{\mu}\!\equiv\!P_{\mu}\label{P}
\end{eqnarray}
so that the spinorial covariant derivative is writable as
\begin{eqnarray}
&\boldsymbol{\nabla}_{\mu}\psi\!=\!(\nabla_{\mu}\ln{\phi}\mathbb{I}
\!-\!\frac{i}{2}\nabla_{\mu}\beta\boldsymbol{\pi}
\!-\!iP_{\mu}\mathbb{I}\!-\!\frac{1}{2}R_{ij\mu}\boldsymbol{\sigma}^{ij})\psi
\label{decspinder}
\end{eqnarray}
from which 
\begin{eqnarray}
&\nabla_{\mu}s_{i}\!=\!R_{ji\mu}s^{j}\label{ds}\\
&\nabla_{\mu}u_{i}\!=\!R_{ji\mu}u^{j}\label{du}
\end{eqnarray}
telling that (\ref{R}, \ref{P}) are true tensors: from the commutator of the covariant derivatives of the spinor or from the commutator of the covariant derivatives of either the velocity or the spin we see that we still have the curvatures
\begin{eqnarray}
&\!\!\!\!\!\!\!\!R^{i}_{\phantom{i}j\mu\nu}\!=\!-(\nabla_{\mu}R^{i}_{\phantom{i}j\nu}
\!-\!\!\nabla_{\nu}R^{i}_{\phantom{i}j\mu}
\!\!+\!R^{i}_{\phantom{i}k\mu}R^{k}_{\phantom{k}j\nu}
\!-\!R^{i}_{\phantom{i}k\nu}R^{k}_{\phantom{k}j\mu})\label{Riemann}\\
\!\!\!\!&qF_{\mu\nu}\!=\!-(\nabla_{\mu}P_{\nu}\!-\!\nabla_{\nu}P_{\mu})\label{Maxwell}
\end{eqnarray}
as it can also be seen with a straightforward substitution, and which tell us that the parameters defined in (\ref{Lorentz}) do not generate any curvature tensor of their own whatever.

Normally $\Omega^{a}_{b\pi}$ and $A_{\mu}$ are the objects inside which all the physical information is stored beside the information regarding the reference system, while $\partial_{\mu}\theta_{ij}$ and $\partial_{\mu}\alpha$ arise from local frame transformations alone and thus with no additional physical information; so $R_{ij\nu}$ and $P_{\mu}$ contain the physical information and an appropriately combined information regarding both the reference system and the local frame: as such (\ref{R}, \ref{P}) have the content of information of connections but as we have also discussed they do follow the transformation law proper of true tensors.

As a consequence in (\ref{R}, \ref{P}) the physical information is now accompanied by a type of information that is entirely due to the frame, and as such susceptible to be vanished by choice, but also truly tensorial, and as such susceptible never to be vanished by choice; the only way out seems to be that this information be always zero, but if this were the case (\ref{R}, \ref{P}) would vanish when no physical force has dynamical effects. As we will see, this does not happen.

\subsection{Dynamic field equations: coupling equations}
For the dynamics, we assume the action given by
\begin{eqnarray}
\nonumber
&\mathscr{L}\!=\!\frac{1}{4}(\partial W)^{2}\!-\!\frac{1}{2}M^{2}W^{2}
\!+\!R\!+\!\frac{1}{4}F^{2}-\\
&-i\overline{\psi}\boldsymbol{\gamma}^{\mu}\boldsymbol{\nabla}_{\mu}\psi
\!+\!XS^{\mu}W_{\mu}\!+\!m\Phi
\label{l}
\end{eqnarray}
in which $(\partial W)_{\mu\nu}$ is the curl of $W_{\mu}$ being the torsion axial vector, and where the generality apparently lost when we defined connections with no torsion is restored by having torsion included as a massive axial vector field \cite{Fabbri:2017fac}.

Varying the above Lagrangian functional with respect to the fields and employing the polar form of the spinor field gives the Dirac matter field equations in polar form
\begin{eqnarray}
\nonumber
&\frac{1}{2}\varepsilon_{\mu\alpha\nu\iota}R^{\alpha\nu\iota}
\!-\!2P^{\iota}u_{[\iota}s_{\mu]}+\\
&+2(\nabla\beta/2\!-\!XW)_{\mu}\!+\!2s_{\mu}m\cos{\beta}\!=\!0\label{dep1}\\
\nonumber
&R_{\mu a}^{\phantom{\mu a}a}
\!-\!2P^{\rho}u^{\nu}s^{\alpha}\varepsilon_{\mu\rho\nu\alpha}+\\
&+2s_{\mu}m\sin{\beta}\!+\!\nabla_{\mu}\ln{\phi^{2}}\!=\!0\label{dep2}
\end{eqnarray}
specifying all the first-order derivatives of the module and the YT angle \cite{Fabbri:2017pwp} as well as the geometric field equations
\begin{eqnarray}
&\nabla^{2}P^{\mu}
\!-\!\nabla_{\sigma}\nabla^{\mu}P^{\sigma}\!=\!-2q^{2}\phi^{2}u^{\mu}\label{me}
\end{eqnarray}
alongside to
\begin{eqnarray}
&\!\!\!\!\nabla^{2}(XW)^{\mu}\!-\!\nabla_{\alpha}\nabla^{\mu}(XW)^{\alpha}
\!+\!M^{2}XW^{\mu}\!=\!2X^{2}\phi^{2}s^{\mu}\label{se}
\end{eqnarray}
as well as
\begin{eqnarray}
\nonumber
&\nabla_{k}R^{ka}_{\phantom{ka}a}g^{\rho\sigma}\!-\!\nabla_{i}R^{i\sigma\rho}
\!-\!\nabla^{\rho}R^{\sigma i}_{\phantom{\sigma i}i}\!+\!R_{ki}^{\phantom{ki}i}R^{k\sigma\rho}+\\
\nonumber
&+R_{ik}^{\phantom{ik}\rho}R^{k\sigma i}
\!-\!\frac{1}{2}R_{ki}^{\phantom{ki}i}R^{ka}_{\phantom{ka}a}g^{\rho\sigma}-\\
\nonumber
&-\frac{1}{2}R^{ika}R_{kai}g^{\rho\sigma}\!=\!\frac{1}{2}[M^{2}(W^{\rho}W^{\sigma}
\!\!-\!\!\frac{1}{2}W^{\alpha}W_{\alpha}g^{\rho\sigma})+\\
\nonumber
&+\frac{1}{4}(\partial W)^{2}g^{\rho\sigma}
\!-\!(\partial W)^{\sigma\alpha}(\partial W)^{\rho}_{\phantom{\rho}\alpha}+\\
\nonumber
&+\frac{1}{4}F^{2}g^{\rho\sigma}\!-\!F^{\rho\alpha}\!F^{\sigma}_{\phantom{\sigma}\alpha}-\\
\nonumber
&-\phi^{2}[(XW\!-\!\nabla\frac{\beta}{2})^{\sigma}s^{\rho}
\!+\!(XW\!-\!\nabla\frac{\beta}{2})^{\rho}s^{\sigma}-\\
\nonumber
&-P^{\sigma}u^{\rho}\!-\!P^{\rho}u^{\sigma}+\\
&+\frac{1}{4}R_{ij}^{\phantom{ij}\sigma}\varepsilon^{\rho ijk}s_{k}
\!+\!\frac{1}{4}R_{ij}^{\phantom{ij}\rho}\varepsilon^{\sigma ijk}s_{k}]]\label{ee}
\end{eqnarray}
specifying second-order derivatives of the gauge potentials, torsion and tetrad fields: these are the field equations that couple electrodynamics, torsion and gravity to the currents, spin and energy densities, respectively \cite{Fabbri:2017xlx}.

As shown above, Riemann and Maxwell tensors can be written in terms of the $R_{ijk}$ and $P_{a}$ tensors, in the same way in which the curl of torsion can be written in terms of the $XW_{a}$ tensor; this means that the electrodynamic and gravitational field equations can be written in terms of the $R_{ijk}$ and $P_{a}$ tensors, once again in the same way in which the torsional field equations are written in terms of the $XW_{a}$ tensor: all geometrical field equations, like the material field equations, are written in terms of tensors.

Apart for the fact that torsion is massive, the symmetry between the torsional and electrodynamic field equations is remarkable; in the gravitational field equations, the energy has a term due to the coupling of electrodynamics and velocity as well as a term due to the coupling of torsion and spin. Curiously there also is a term due to the coupling between gravity itself and again the spin.
\section{The Hydrogen Atom}
In the preliminary section we have seen that for a Dirac spinor field we can establish the existence of objects behaving like connections but being true tensors, specifying that as such they may be thought to vanish once physical interactions are absent, and eventually warning readers that this does not happen: to see that we can not have the vanishing of this \emph{tensorial connection} we furnish one notable example of exact solution for the field equations.

For the system we want to study, we neglect torsion-gravity focusing only on an electrodynamic field, which is taken as external and non-dynamic with structure
\begin{eqnarray}
&qA_{t}\!=\!-\alpha/r
\end{eqnarray}
where $q^{2}\!\equiv\!\alpha$ being the fine-structure constant as usual.

\subsection{The standard treatment}
We begin with the standard treatment of the hydrogen atom, specializing to the $1S$ orbital of minimal energy.

It is possible to look for solutions for which the entire temporal dependence is taken into account by the energy eigen-state $i\partial_{t}\psi\!=\!E\psi$ so that in spherical coordinates 
\begin{eqnarray}
&\vec{r}\!=\!\left(\begin{array}{c}
\!r\sin{\theta}\cos{\varphi}\!\\
\!r\sin{\theta}\sin{\varphi}\!\\
\!r\cos{\theta}\!
\end{array}\right)
\end{eqnarray}
the field equations are usually written according to
\begin{eqnarray}
\nonumber
&(E\!+\!\frac{\alpha}{r})\!\left(\begin{array}{cc}
\!\mathbb{I} & \ \ 0 \\
\!0 & -\mathbb{I}\!
\end{array}\right)\!\psi
\!+\!\frac{i}{r}\!\left(\begin{array}{cc}
\!0 & \ \vec{\boldsymbol{\sigma}}\!\cdot\!\vec{r}\ \\
\!-\vec{\boldsymbol{\sigma}}\!\cdot\!\vec{r}\ & 0\!
\end{array}\right)\!\partial_{r}\psi-\\
&-\frac{i}{r^{2}}\!\left(\begin{array}{cc}
\!0 & \ \vec{\boldsymbol{\sigma}}\!\cdot\!\vec{r}\ \vec{\boldsymbol{\sigma}}\!\cdot\!\vec{L}\ \\
\!-\vec{\boldsymbol{\sigma}}\!\cdot\!\vec{r}\ \vec{\boldsymbol{\sigma}}\!\cdot\!\vec{L}\ & 0\!
\end{array}\right)\!\psi\!-\!m\psi\!=\!0\label{dee}
\end{eqnarray}
as it is easy to check in common textbooks; it is clear that in order to look for solutions given in terms of separation of variables, this form is the most apt because the radial dependence is separated from the angular dependence as it can be seen from the fact that the angular momentum
\begin{eqnarray}
&\vec{L}F\!=\!\left(\begin{array}{c}
\!i\sin{\varphi}\partial_{\theta}F\!+\!i\cot{\theta}\cos{\varphi}\partial_{\varphi}F\!\\
\!-i\cos{\varphi}\partial_{\theta}F\!+\!i\cot{\theta}\sin{\varphi}\partial_{\varphi}F\!\\
\!-i\partial_{\varphi}F\!
\end{array}\right)\label{ang}
\end{eqnarray}
for any function $F$ is given in terms of the elevation and the azimuthal angles alone: so, solutions for eigen-states of energy after variables separation will additionally be for eigen-states of angular momentum, and consequently they can be classified in terms of these quantum numbers.

For example, the minimum of the energy states corresponds to the $1S$ orbital given with $E\!=\!m\Gamma$ in the form
\begin{eqnarray}
&\psi\!=\!\frac{1}{\sqrt{1+\Gamma}}r^{\Gamma-1}e^{-\alpha m r}e^{-im\Gamma t}\left(\begin{array}{c}
\!1\!+\!\Gamma\!\\
\!0\!\\
\!i\alpha\cos{\theta}\!\\
\!i\alpha\sin{\theta}e^{i\varphi}\!
\end{array}\right)\label{solution}
\end{eqnarray}
where $\Gamma\!=\!\sqrt{1-\alpha^{2}}$ as it can be seen by plugging it into equations (\ref{dee}) with operator (\ref{ang}) and verifying that the computations check out. It is then straightforward to see that, introducing $\Delta(\theta)\!=\!1/\!\sqrt{1\!-\!\alpha^{2}|\!\sin{\theta}|^{2}}$ for the sake of simplicity, the vectors are given according to expressions
\begin{eqnarray}
&s^{0}\!=\!0\label{s0}\\
&s^{1}\!=\!(1\!-\!\Gamma)\Delta(\theta)\sin{\theta}\cos{\theta}\cos{\varphi}\label{s1}\\
&s^{2}\!=\!(1\!-\!\Gamma)\Delta(\theta)\sin{\theta}\cos{\theta}\sin{\varphi}\label{s2}\\
&s^{3}\!=\!\Delta(\theta)(|\!\cos{\theta}|^{2}\!+\!\Gamma|\!\sin{\theta}|^{2})\label{s3}
\end{eqnarray}
\begin{eqnarray}
&u^{0}\!=\!\Delta(\theta)\label{u0}\\
&u^{1}\!=\!-\alpha\Delta(\theta)\sin{\theta}\sin{\varphi}\label{u1}\\
&u^{2}\!=\!\alpha\Delta(\theta)\sin{\theta}\cos{\varphi}\label{u2}\\
&u^{3}\!=\!0\label{u3}
\end{eqnarray}
while
\begin{eqnarray}
&\beta\!=\!-\arctan{(\frac{\alpha}{\Gamma}\cos{\theta})}\label{b}
\end{eqnarray}
\begin{eqnarray}
&\phi\!=\!r^{\Gamma-1}e^{-\alpha m r}(1-\alpha^{2}|\!\sin{\theta}|^{2})^{\frac{1}{4}}\label{m}
\end{eqnarray}
showing that spin and velocity are orthogonal while the velocity does not have the third component and that the YT angle depends only on the elevation angle while the module depends on both radial coordinate and elevation angle, thus making clear that in relativistic cases also the minimum energy $1S$ orbital is not spherically symmetric.

We will now move to a treatment less standard.

\subsection{The covariant approach}
Equations (\ref{dee}) with (\ref{ang}) are standard, but it is not at all clear how they are a specific case of the most general Dirac spinor field equations written in terms of covariant derivatives: next we are going to render this manifest by showing how (\ref{dee}, \ref{ang}) are the Dirac equation written with a spinorial covariant derivative in a very special case.

In spherical coordinates the metric tensor is given by
\begin{eqnarray}
&g_{tt}\!=\!1\\
&g_{rr}\!=\!-1\\
&g_{\theta\theta}\!=\!-r^{2}\\
&g_{\varphi\varphi}\!=\!-r^{2}|\!\sin{\theta}|^{2}
\end{eqnarray}
with connection
\begin{eqnarray}
&\Lambda^{\theta}_{\theta r}\!=\!\frac{1}{r}\\
&\Lambda^{r}_{\theta\theta}\!=\!-r\\
&\Lambda^{\varphi}_{\varphi r}\!=\!\frac{1}{r}\\
&\Lambda^{r}_{\varphi\varphi}\!=\!-r|\!\sin{\theta}|^{2}\\
&\Lambda^{\varphi}_{\varphi\theta}\!=\!\cot{\theta}\\
&\Lambda^{\theta}_{\varphi\varphi}\!=\!-\cot{\theta}|\!\sin{\theta}|^{2}
\end{eqnarray}
giving a Riemann curvature identically equal to zero as expected; when treating spinors it is necessary to introduce tetrad vectors, and although their choice is unique only up to Lorentz transformations, the tetrad vectors
\begin{eqnarray}
&\!\!\!\!e^{0}_{t}\!=\!1\\
&\!\!\!\!e^{1}_{r}\!=\!\sin{\theta}\cos{\varphi}\ \ \ \ 
e^{2}_{r}\!=\!\sin{\theta}\sin{\varphi}\ \ \ \ 
e^{3}_{r}\!=\!\cos{\theta}\\
&\!\!\!\!e^{1}_{\theta}\!=\!r\cos{\theta}\cos{\varphi}\ \ \ 
e^{2}_{\theta}\!=\!r\cos{\theta}\sin{\varphi}\ \ \
e^{3}_{\theta}\!=\!-r\sin{\theta}\\
&\!\!\!\!e^{1}_{\varphi}\!=\!-r\sin{\theta}\sin{\varphi}\ \ \ \ 
e^{2}_{\varphi}\!=\!r\sin{\theta}\cos{\varphi}
\end{eqnarray}
and
\begin{eqnarray}
&\!\!\!\!e_{0}^{t}\!=\!1\\
&\!\!\!\!e_{1}^{r}\!=\!\sin{\theta}\cos{\varphi}\ \ \ \ 
e_{2}^{r}\!=\!\sin{\theta}\sin{\varphi}\ \ \ \ 
e_{3}^{r}\!=\!\cos{\theta}\\
&\!\!\!\!e_{1}^{\theta}\!=\!\frac{1}{r}\cos{\theta}\cos{\varphi}\ \ \ 
e_{2}^{\theta}\!=\!\frac{1}{r}\cos{\theta}\sin{\varphi}\ \ \
e_{3}^{\theta}\!=\!-\frac{1}{r}\sin{\theta}\\
&\!\!\!\!e_{1}^{\varphi}\!=\!-\frac{1}{r\sin{\theta}}\sin{\varphi}\ \ \ \ 
e_{2}^{\varphi}\!=\!\frac{1}{r\sin{\theta}}\cos{\varphi}
\end{eqnarray}
constitute a special choice since they are those whose spin connection identically vanishes: this is the special choice of tetrad vectors giving rise to the covariant derivative in terms of which the Dirac spinor field equation is identical to what we obtain when (\ref{ang}) is plugged into (\ref{dee}) above, and so we did manage to write the standard form of Dirac equation as a special case of the general Dirac equation.

From (\ref{ds}, \ref{du}) and (\ref{s0}--\ref{u3}) we can see that
\begin{eqnarray}
&\partial_{\theta}\theta_{01}\!=\!\alpha\cos{\theta}|\Delta(\theta)|^{2}\sin{\varphi}\\
&\partial_{\theta}\theta_{02}\!=\!-\alpha\cos{\theta}|\Delta(\theta)|^{2}\cos{\varphi}\\
&\partial_{\theta}\theta_{23}\!=\!(1\!-\!\Gamma|\Delta(\theta)|^{2})\sin{\varphi}\\
&\partial_{\theta}\theta_{31}\!=\!-(1\!-\!\Gamma|\Delta(\theta)|^{2})\cos{\varphi}\\
&\partial_{\varphi}\theta_{12}\!=\!-1
\end{eqnarray}
in this frame: since the spin connection vanishes then we have that $\partial_{\mu}\theta_{ij}
\!\equiv\!R_{ij\mu}$ and these constitute a true tensor.

And additionally, we have that $P_{t}\!=\!E+\alpha/r$ identically.

With these components of $R_{ij\mu}$ and $P_{\mu}$ and considering expressions of $s^{a}$ and $u^{a}$ and (\ref{b}, \ref{m}) then (\ref{dep1}, \ref{dep2}) hold.

We will now employ covariance to have this special case converted into a complementary type of special case.

\subsection{The polar form}
In the previous subsection, we studied the polar form in the very special case where $\Omega_{ij\mu}\!=\!0$ identically; on the other hand, we can take advantage of covariance to move into the system of reference in which the polar form is in the complementary case where $\partial_{\mu}\theta_{ij}\!=\!0$ identically: this case corresponds to a spinor at rest and with spin aligned along the third axis and with the positive orientation.

We employ the possibility to rotate the spin along the third axis and boost into the rest frame so that
\begin{eqnarray}
&\psi\!=\!\phi\ e^{-i(m\Gamma t-\frac{1}{2}\varphi)}
e^{-\frac{i}{2}\beta\boldsymbol{\pi}}
\!\left(\begin{array}{c}
\!\!\sqrt{2}\!\\
\!\!0\!\\
\!\!0\!\\
\!\!0\!
\end{array}\right)
\end{eqnarray}
which is the polar form; then the spin has only the third spatial component whereas the velocity has only the time component and we still have the degrees of freedom
\begin{eqnarray}
&\beta\!=\!-\arctan{(\frac{\alpha}{\Gamma}\cos{\theta})}
\end{eqnarray}
\begin{eqnarray}
&\phi\!=\!r^{\Gamma-1}e^{-\alpha m r}(1-\alpha^{2}|\!\sin{\theta}|^{2})^{\frac{1}{4}}
\end{eqnarray}
as it should have been expected for they are scalar fields.

Then the metric tensor and the connection do not have any modification; the tetrad vectors become
\begin{eqnarray}
&\!\!\!\!e^{0}_{t}\!=\!\cosh{(2\eta)}\ \ \ \ 
e^{2}_{t}\!=\!-\sinh{(2\eta)}\\
&\!\!\!\!e^{1}_{r}\!=\!\sin{(\theta\!-\!2\zeta)}\ \ \ \ 
e^{3}_{r}\!=\!\cos{(\theta\!-\!2\zeta)}\\
&\!\!\!\!e^{1}_{\theta}\!=\!r\cos{(\theta\!-\!2\zeta)}\ \ \ \ 
e^{3}_{\theta}\!=\!-r\sin{(\theta\!-\!2\zeta)}\\
&\!\!\!\!e^{0}_{\varphi}\!=\!-r\sin{\theta}\sinh{(2\eta)}\ \ \ \ 
e^{2}_{\varphi}\!=\!r\sin{\theta}\cosh{(2\eta)}
\end{eqnarray}
and
\begin{eqnarray}
&\!\!\!\!e_{0}^{t}\!=\!\cosh{(2\eta)}\ \ \ \ 
e_{2}^{t}\!=\!\sinh{(2\eta)}\\
&\!\!\!\!e_{1}^{r}\!=\!\sin{(\theta\!-\!2\zeta)}\ \ \ \ 
e_{3}^{r}\!=\!\cos{(\theta\!-\!2\zeta)}\\
&\!\!\!\!e_{1}^{\theta}\!=\!\frac{1}{r}\cos{(\theta\!-\!2\zeta)}\ \ \ \ 
e_{3}^{\theta}\!=\!-\frac{1}{r}\sin{(\theta\!-\!2\zeta)}\\
&\!\!\!\!e_{0}^{\varphi}\!=\!\frac{1}{r\sin{\theta}}\sinh{(2\eta)}\ \ \ \ 
e_{2}^{\varphi}\!=\!\frac{1}{r\sin{\theta}}\cosh{(2\eta)}
\end{eqnarray}
where we have the identities
\begin{eqnarray}
&\sin{(\theta\!-\!2\zeta)}\!=\!\Gamma\sin{\theta}\Delta\\
&\cos{(\theta\!-\!2\zeta)}\!=\!\cos{\theta}\Delta
\end{eqnarray}
and
\begin{eqnarray}
&\sinh{(2\eta)}\!=\!\alpha\sin{\theta}\Delta\\
&\cosh{(2\eta)}\!=\!\Delta
\end{eqnarray}
with
\begin{eqnarray}
&\sin{(2\zeta)}\!=\!(1\!-\!\Gamma)\cos{\theta}\sin{\theta}\Delta\\
&\cos{(2\zeta)}\!=\!(|\!\cos{\theta}|^{2}+\Gamma|\!\sin{\theta}|^{2})\Delta
\end{eqnarray}
as well as
\begin{eqnarray}
&\sinh{\eta}\!=\!\frac{1}{\sqrt{2}}\sqrt{\Delta\!-\!1}\\
&\cosh{\eta}\!=\!\frac{1}{\sqrt{2}}\sqrt{\Delta\!+\!1}
\end{eqnarray}
and
\begin{eqnarray}
&\sin{\zeta}\!=\!\frac{1}{\sqrt{2}}
\sqrt{1\!-\!(|\!\cos{\theta}|^{2}\!+\!\Gamma|\!\sin{\theta}|^{2})\Delta}\\
&\cos{\zeta}\!=\!\frac{1}{\sqrt{2}}
\sqrt{1\!+\!(|\!\cos{\theta}|^{2}\!+\!\Gamma|\!\sin{\theta}|^{2})\Delta}
\end{eqnarray}
in terms of the angle
\begin{eqnarray}
\zeta\!=\!\arctan{\left(\!\sqrt{\frac{1\!-\!(|\!\cos{\theta}|^{2}
\!+\!\Gamma|\!\sin{\theta}|^{2})\Delta} {1\!+\!(|\!\cos{\theta}|^{2}
\!+\!\Gamma|\!\sin{\theta}|^{2})\Delta}}\ \right)}
\end{eqnarray}
and the rapidity
\begin{eqnarray}
\eta\!=\!\ln{\left(\!\sqrt{(\Delta\!+\!1)/2}+\sqrt{(\Delta\!-\!1)/2}\ \right)}
\end{eqnarray}
of the rotation and the boost that are needed in order to obtain the spinor field in the full polar form.

As a consequence, we have the expressions
\begin{eqnarray}
&R_{t\varphi\theta}\!=\!-\alpha r \sin{\theta}\cos{\theta}|\Delta(\theta)|^{2}\\
&R_{r\theta\theta}\!=\!-r(1\!-\!\Gamma|\Delta(\theta)|^{2})\\
&R_{r\varphi\varphi}\!=\!-r|\!\sin{\theta}|^{2}\\
&R_{\theta\varphi\varphi}\!=\!-r^{2}\sin{\theta}\cos{\theta}
\end{eqnarray}
as well as $P_{t}\!=\!E\!+\!\alpha/r$ and $P_{\varphi}\!=\!-1/2$ identically.

With all these elements it is now possible to check that the Dirac spinor field equations are verified indeed.

We may now proceed to discuss all these elements.
\section{Tensorial connections}
In the first section we introduced the $R_{ijk}$ and $P_{a}$ coefficients discussing how they behave like connections but being true tensors. In the second section we studied the hydrogen atom as an example of a system where we could actually compute these objects in an explicit manner.

Now we will employ this prototypical case to extract information about the $R_{ijk}$ and $P_{a}$ tensorial connections.

Despite employing a very particular situation we will still try to gather information of general validity.

\subsection{Invariant combinations}
The coefficients $R_{ijk}$ and $P_{a}$ are real tensor fields, and consequently it makes sense to decompose them in irreducible parts: $P_{a}$ is irreducible but $R_{ijk}$ has trace
\begin{eqnarray}
&\!R_{a}\!=\!R_{ac}^{\phantom{ac}c}
\end{eqnarray}
the dual of its completely antisymmetric part 
\begin{eqnarray}
&\!\!B_{a}\!=\!\frac{1}{2}\varepsilon_{aijk}R^{ijk}
\end{eqnarray}
and with which it is possible to obtain the non-completely antisymmetric irreducible part given by the expression
\begin{eqnarray}
&\!\!\!\!\Pi_{ijk}\!=\!R_{ijk}\!-\!\frac{1}{3}(R_{i}\eta_{jk}\!-\!R_{j}\eta_{ik})
\!-\!\frac{1}{3}\varepsilon_{ijka}B^{a}
\end{eqnarray}
such that $\Pi_{ia}^{\phantom{ia}a}\!=\!0$ and $\Pi_{ijk}\varepsilon^{ijka}\!=\!0$ hold identically.

We can also compute the invariants, the simplest being given by the squared tensors: after a simple inventory we see that from $P_{a}$ we can compute only $P_{a}P^{a}$ whereas for the $R_{ijk} \!=\! -R_{jik}$ we have the possible contractions given by the scalars $R_{ac}^{\phantom{ac}c} R^{ai}_{\phantom{ai}i}$ and $\!R_{ijk}R^{kij}$ with $\!R_{ijk}R^{ijk}$ as well as the pseudo-scalars $R_{pq}^{\phantom{pq}q}R_{ijk}\varepsilon^{pijk}$ and $\!R_{ijc}R_{pq}^{\phantom{pq}c}\varepsilon^{ijpq}$ or
\begin{eqnarray}
&R_{ac}^{\phantom{ac}c}R^{ai}_{\phantom{ai}i}\!=\!R_{a}R^{a}
\end{eqnarray}
and
\begin{eqnarray}
&R_{ijk}R^{kij}\!=\!-\frac{1}{3}R_{i}R^{i}\!-\!\frac{2}{3}B_{a}B^{a}
\!-\!\frac{1}{2}\Pi_{ijk}\Pi^{ijk}
\end{eqnarray}
with
\begin{eqnarray}
&\frac{1}{2}R_{ijk}R^{ijk}\!=\!\frac{1}{3}R_{i}R^{i}\!-\!\frac{1}{3}B_{a}B^{a}
\!+\!\frac{1}{2}\Pi_{ijk}\Pi^{ijk}
\end{eqnarray}
as well as
\begin{eqnarray}
&\frac{1}{2}R_{pq}^{\phantom{pq}q}R_{ijk}\varepsilon^{pijk}\!=\!R_{a}B^{a}
\end{eqnarray}
and
\begin{eqnarray}
&\frac{1}{4}R_{ijc}R_{pq}^{\phantom{pq}c}\varepsilon^{ijpq}\!=\!\frac{2}{3}R_{a}B^{a}
\!+\!\frac{1}{4}\Pi_{ijc}\Pi_{pq}^{\phantom{pq}c}\varepsilon^{ijpq}
\end{eqnarray}
written in terms of the irreducible parts and thus showing that these five are all the independent contractions.

If we consider field equations (\ref{dep1}, \ref{dep2}), it also becomes possible to establish a link between the $P_{a}$ and $R_{ijk}$ as
\begin{eqnarray}
\nonumber
&P^{\mu}\!=\!m\cos{\beta}u^{\mu}
\!-\!\frac{1}{2}(\nabla_{k}\beta\!-\!2XW_{k}\!+\!B_{k})s^{[k}u^{\mu]}-\\
&-\frac{1}{2}(\nabla_{k}\ln{\phi^{2}}\!+\!R_{k})s_{j}u_{i}\varepsilon^{kji\mu}
\end{eqnarray}
although it is only a link between $P_{a}$ and the two vectorial parts of $R_{ijk}$ and not a link to the full tensor.

By evaluating the square we obtain the relationships
\begin{eqnarray}
\nonumber
&P_{a}P^{a}\!=\!m^{2}|\!\cos{\beta}|^{2}
\!-\!m\cos{\beta}(\nabla\beta\!-\!2XW\!+\!B)\!\cdot\!s+\\
\nonumber
&+\left|\frac{1}{2}(\nabla\beta\!-\!2XW\!+\!B)\!\cdot\!s\right|^{2}
\!-\!\left|\frac{1}{2}(\nabla\beta\!-\!2XW\!+\!B)\!\cdot\!u\right|^{2}+\\
\nonumber
&+\left|\frac{1}{2}(\nabla\ln{\phi^{2}}\!+\!R)\!\cdot\!s\right|^{2}
\!-\!\left|\frac{1}{2}(\nabla\ln{\phi^{2}}\!+\!R)\!\cdot\!u\right|^{2}+\\
&+\frac{1}{4}|\nabla\ln{\phi^{2}}\!+\!R|^{2}
\end{eqnarray}
which are the relationships of dispersion in general cases.

\subsection{Special cases}
The actual computation in the special case that we are considering furnishes the following six combinations
\begin{eqnarray}
&\!P_{a}P^{a}\!=\!(m\Gamma\!+\!\alpha/r)^{2}\!-\!\left|\frac{1}{2r\sin{\theta}}\right|^{2}\\
&\!\!R_{a}R^{a}\!=\!-\frac{1}{r^{2}}\!
\left[(2\!-\!\Gamma\Delta^{2})^{2}\!+\!|\!\cot{\theta}|^{2}\right]\\
&\!\!\!\!B_{a}B^{a}\!=\!-\frac{1}{r^{2}}\alpha^{2}|\!\cos{\theta}|^{2}\Delta^{4}\\
&\!\!\!\!\!\!\!\!\frac{1}{2}R_{ijk}R^{ijk}\!=\!\frac{1}{r^{2}}\!
\left[\alpha^{2}|\!\cos{\theta}|^{2}\Delta^{4}\!-\!(1\!-\!\Gamma\Delta^{2})^{2}
\!-\!\frac{1}{|\!\sin{\theta}|^{2}}\right]\\
&\!\!R_{a}B^{a}\!=\!\frac{1}{r^{2}}\alpha\cos{\theta}\Delta^{2}(2\!-\!\Gamma\Delta^{2})\\
&\!\!\!\!\frac{1}{4}R_{ijc}R_{pq}^{\phantom{pq}c}\varepsilon^{ijpq}
\!=\!\frac{2}{r^{2}}\alpha\cos{\theta}\Delta^{2}(1\!-\!\Gamma\Delta^{2})
\end{eqnarray}
as the total number of independent squared scalar fields.

If the hydrogen atom is taken when the fine structure constant goes to zero we obtain the free case as
\begin{eqnarray}
&P_{a}P^{a}\!=\!m^{2}\!+\!\frac{1}{4}R_{a}R^{a}
\!=\!m^{2}\!-\!\left|\frac{1}{2r\sin{\theta}}\right|^{2}
\end{eqnarray}
showing that not even in free case are these tensors zero.

Hence we have a situation where some of the independent squared invariants built with $P_{a}$ and $R_{ijk}$ do remain different from zero even in the free field configuration.
\section{Comment}
In the first section we have seen that $R_{ijk}$ and $P_{a}$ have the character of connections while being true tensors; and in the second section we have seen examples showing that they are not zero: thus they are generally non-vanishing tensors behaving like connections, and as a consequence we found it natural to call them tensorial connections.

Such tensorial connections, because of the way in which they are constructed, contain both information about the physical interactions and the reference system, but since they are tensors, this information is to be covariant under frame transformations: the information about a physical interaction is always covariant under frame transformations, but the information about the reference system is in general not, so we wondered whether it could be zero in any circumstance, although we found one notable situation in which it was not equal to zero whatsoever.

As a consequence, we must admit the existence of types of information that despite being pertinent to a reference system, nevertheless are frame independent establishing the existence of truly covariant inertial forces.

These forces are still inertial in their lack of coupling to sources, since they possess no curvature.

\subsection{Interpretation}
Even if there is no coupling for these inertial forces, the fact that these forces are at a time inertial and covariant poses some problem of meaning because they do not look like anything we already know. On the other hand, some analogy may be formed if we consider that they seem to appear only in spinorial systems and so something must be related to the spin structure. A parallel can be made on analogies between orbital angular momentum and spin in the case of the macroscopic and the spinorial systems.

For macroscopic systems, the point-like particle has an orbital angular momentum, but we can always render it equal to zero into the particle rest-frame. Instead for the spinor systems, the spin behaves as some orbital angular momentum, but it is different for it can never vanish.

As for macroscopic systems, both orbital angular momenta and inertial forces are present for rotational motions, but they can always be removed in the rest-frame, similarly, for spinorial systems, both spin and covariant inertial forces are present, and they are truly covariant.

Such \emph{covariant inertial forces} and the spin are covariant because they are intrinsic features of spinor fields.

Not because they are effects emerging from motion.
\section{Conclusion}
In this paper we have seen that for spinorial fields, it is possible to define two objects $P_{a}$ and $R_{ijk}$ which have the apparently contradictory feature of containing all information of a connection, that is both the information about physical interactions and about the reference system, but being tensors, that is frame-independent; physical interactions are known to be covariant, while information about the reference system is in general not, although we have seen that this type of information about the reference system but being frame-independent cannot be simply dismissed as inexistent, since we gave an example of situation in which it was not: therefore we were forced to assume the existence of generally non-vanishing tensors behaving like connections, which we called tensorial connections, encoding a type of information that can be thought as a covariant inertial force. These forces are still inertial in their lacking the coupling to a source.

In their appearing only for spinorial systems, and being objects related to rotations while being covariant, these covariant inertial forces parallel the spin: both are intrinsic features. And neither results from a motion.

So far as we know, this is the first occurrence in which such covariant inertial forces for the spinorial field is discussed anywhere in existing literature.

\end{document}